\magnification1200
\parindent0pt
\parskip3pt
\nopagenumbers
\newdimen\interlinia
\interlinia=\baselineskip
\baselineskip=1.5\baselineskip
\input amssym.tex

\font\kapitaliki=plcsc10

\def\startcenter{%
  \par
  \begingroup
  \leftskip=0pt plus 1fil
  \rightskip=\leftskip
  \parindent=0pt
  \parfillskip=0pt
}
\def\stopcenter{%
  \par
  \endgroup
}
\long\def\centerpars#1{\startcenter#1\stopcenter}

\def\roz#1{\bigskip\centerpars{\kapitaliki #1}\bigskip}

\def\dzien{\sl \number\day/\number\month/\number\year}
\footline{\hss{\sl \ \dzien;\ \folio}}

\footline={\hss {\sl Page\ \folio}}


\centerline{\kapitaliki Confidence Interval for Quantile Ratio of the Dagum Distribution}

\vskip2truecm
\centerline{\kapitaliki Alina Jêdrzejczak$^1$, Dorota Pekasiewicz$^1$, Wojciech Zieliñski$^{2}$}

\bigskip

\startcenter
\baselineskip=\interlinia
$^1$ Institute of Statistics and Demography\par
University of Lodz\par
Rewolucji 1905 r. 41/43, PL-90-214 £ódŸ\par
e-mail: jedrzej@uni.lodz.pl\par
e-mail: pekasiewicz@uni.lodz.pl

\bigskip

$^2$ Department of Econometrics and Statistics\par
Warsaw University of Life Scinces\par
Nowoursynowska 159, PL-02-787 Warszawa\par
e-mail: wojciech$\_$zielinski@sggw.pl\par
http://wojtek.zielinski.statystyka.info

\stopcenter

\vskip2truecm

\bigskip

\begingroup
\leftskip2truecm
\rightskip=\leftskip
\parindent=0pt

In economic research inequality measures based on ratios of quantiles are frequently applied to the analysis of income distributions. In the paper, we construct a confidence interval for such measures under
the Dagum distribution which has been widely assumed as a~model for income distributions in empirical analyses Its properties are investigated on the basis of computer simulations. The constructed confidence interval is applied to the analysis of inequality income in Poland in 2015.

\bigskip

{\kapitaliki keywords}: ratio of quantiles, confidence interval, Dagum distribution, quintile share ratio

62F25 62P20

\endgroup

\vfill\eject
\bigskip

\roz{1. introduction}

In Eurostat regional yearbook  (2016) a measure of income distribution inequality is defined as the income quintile share ratio or the $S80/S20$ ratio. It is calculated as the ratio of total income received by the $20\%$ of the population with the highest income (the top quintile) to that received by the $20\%$ of the population with the lowest income (the bottom quintile), i.e. income quintile share ratio is defined as
$$r_{0.2,0.8}={F^{-1}(0.8)\over F^{-1}(0.2)},$$
where $F$ denotes the distribution of the population income. The natural estimator of $r_{0.2,0.8}$ is the ratio of appropriate sample quintiles. However, the problem is in interval estimation. According to the best knowledge of the Authors such problem was never considered. In the paper a confidence interval for the population ratio of quintiles is constructed. The proposed confidence interval is based on the asymptotic distribution of the ratio of sample quintiles.

We confine ourselves to Dagum (1977) distribution as a probabilistic model for income. Dagum distribution is widely used in modeling income in many countries all over the world (see for example Domañski and Jêdrzejczak 1998, Jêdrzejczak 1994). The Dagum distribution has many good mathematical as well as statistical properties. Basic properties of that distribution are presented in Appendix 1; for more see Kleiber (2008), Dey et al. (2017). See also Encyclopedia (2006) (pp. 3363-3378, also 3236-3248) and the references therein.

The paper is organized as follows. In the second section confidence interval for a ratio of quantiles is constructed. It is based on the ratio of sample quantiles of the Dagum distribution. It appears that ends of the proposed confidence interval depend on a shape parameter which should be estimated from a sample. In the third section a short simulation study is provided. In that study two estimators of the shape parameter were applied. Namely, the estimator obtained by the method of moments and the one obtained by the  method of probability-weighted moments. Results of the simulations are very similar for those two estimators. In the fourth section an application to income inequality analysis based on the data coming from the Polish Household Budget Survey is presented. In the last section some conclusions are presented as well as some remarks on further research on the subject.

We consider a more general set-up, namely a confidence interval for a ratio of $\alpha$ and $\beta$ quantiles is constructed. To obtain a confidence interval for a quintile ratio it is enough to put $\alpha=0.2$ and $\beta=0.8$. Results of the paper may be easily generalized to other distributions applied in modeling personal income, such as Pareto, Burr, Beta etc.

\goodbreak
\roz{2. confidence interval}

Let $0<\alpha<\beta<1$ be given numbers and let $$r_{\alpha,\beta}={F^{-1}(\beta)\over F^{-1}(\alpha)},$$
where $F(\cdot)$ is cumulative distribution function (CDF) of the income distribution be the quantile ratio of interest. Let $X_1,\ldots,X_n$ be a sample of incomes of randomly drawn $n$ persons. Let $X_{1:n}\leq\cdots\leq X_{n:n}$ denote the ordered sample. As an estimator of $r_{\alpha,\beta}$ it is taken
$$r^*_{\alpha,\beta}={X_{\lfloor n\beta\rfloor+1:n}\over X_{\lfloor n\alpha\rfloor+1:n}},$$
where $\lfloor x\rfloor$ denotes the greatest integer not greater than $x$.

In our considerations we confine ourselves to the Dagum distribution, i.e. throughout the paper it will be assumed that the distribution of the population income is the Dagum one. As it was mentioned above, the Dagum distribution fits population income quite well for many countries all over the world.

Consider the Dagum distribution with parameters $a>0$, $v>0$ and $\lambda>0$. Its cumulative distribution function (CDF) and probability density function (PDF) are as follows
$$F_{a, v, \lambda}(x)=\left(1+\left({x\over\lambda}\right)^{-v}\right)^{-a}\hbox{ for }x>0$$
and
$$f_{a, v, \lambda}(x)={av\over\lambda}\left({x\over\lambda}\right)^{av-1}\left(1+\left({x\over\lambda}\right)^v\right)^{-a-1}\hbox{ for }x>0.$$
Its quantile function equals
$$Q_{a, v, \lambda}(q)=\lambda\left(q^{-1/a}-1 \right)^{-1/v}\hbox{ for }0<q<1.$$

For other interesting properties of the Dagum distribution see Appendix 1.

The problem is in constructing a confidence interval  at the confidence level $\delta$ for a ratio of quantiles of the Dagum distribution
$$r_{\alpha,\beta}={Q_{a, v, \lambda}(\beta)\over Q_{a, v, \lambda}(\alpha)}=\left({\beta^{-1/a}-1\over\alpha^{-1/a}-1} \right)^{-1/v}$$
on the basis of a random sample $X_1,\ldots,X_n$.

In what follows ``large'' sample sizes are considered, i.e. it is assumed that $n\to\infty$. There are two reasons for such approach. The first one is such that real sample sizes usually comprise many thousands of observations. The second one is rather technical one. Namely, the finite sample size distribution of the ratio of sample quantiles of the Dagum distribution is analytically untractable (for exact distribution see Maswadah 2013).

\bigskip
{\bf Theorem 1.} For $0<\alpha<\beta<1$ the random variable $r^*_{\alpha,\beta}$ is strongly consistent estimator of $r_{\alpha,\beta}$, for all $a,v,\lambda$.
\bigskip

{\sl Proof.} The proof follows form the fact (David and Nagaraja 2003; Serfling 1980) that $X_{\lfloor n\alpha\rfloor+1:n}$ is strongly consistent estimator of the $\alpha$'s quantile of the underlying distribution. Application of Slutsky theorem gives the thesis. $\blacksquare$

\bigskip
{\bf Theorem 2.} For $0<\alpha<\beta<1$ the estimator $r^*_{\alpha,\beta}$ is asymptotically normally distributed random variable.
\bigskip

{\sl Proof.} Let $Y_i=\ln X_i$. Of course $Y_{i:n}=\ln X_{i:n}$. Let $\gamma^Y_\alpha$ and $\gamma^Y_\beta$ denote the quantiles of $Y$. For $\alpha<\beta$ we have (Serfling 1980, th. 2.3.3; David and Nagaraja 2003, th. 10.3):
$$\sqrt{n}\left[\matrix{Y_{\lfloor n\alpha\rfloor+1:n}-\gamma^Y_\alpha\cr Y_{\lfloor n\beta\rfloor+1:n}-\gamma^Y_\beta\cr }\right]\to
N_2\left(\left[\matrix{0\cr0}\right],
\left[\matrix{{\alpha(1-\alpha)\over\left(f_Y(\gamma^Y_\alpha)\right)^2}&{\alpha(1-\beta)\over\left(f_Y(\gamma^Y_\alpha)f_Y(\gamma^Y_\beta)\right)}\cr
{\alpha(1-\beta)\over\left(f_Y(\gamma^Y_\alpha)f_Y(\gamma^Y_\beta)\right)}&{\beta(1-\beta)\over\left(f_Y(\gamma^Y_\beta)\right)^2}\cr}\right]\right),$$
where $f_Y(\cdot)$ is the PDF of $Y$.

Hence
$$\sqrt{n}\left[\left(Y_{\lfloor n\beta\rfloor+1:n}-Y_{\lfloor n\alpha\rfloor+1:n}\right)-\left(\gamma^Y_\beta-\gamma^Y_\alpha\right)\right]\to
N\left(0,\sigma^2\right),$$
where
$$\sigma^2=
{\beta(1-\beta)\over\left(f_Y(\gamma^Y_\beta)\right)^2}+
{\alpha(1-\alpha)\over\left(f_Y(\gamma^Y_{\alpha\phantom{\beta}})\right)^2}-
2{\alpha(1-\beta)\over\left(f_Y(\gamma^Y_\beta)f_Y(\gamma^Y_\alpha)\right)}.$$
So we have
$$\sqrt{n}\left(\ln{X_{\lfloor n\beta\rfloor+1:n}\over X_{\lfloor n\alpha\rfloor+1:n}}-\left(\gamma^Y_\beta-\gamma^Y_\alpha\right)\right)\to N\left(0,\sigma^2\right).$$
Applying Delta method (Greene 2003, p. 913) with $g(t)=e^t$:
$$\sqrt{n}\left({X_{\lfloor n\beta\rfloor+1:n}\over X_{\lfloor n\alpha\rfloor+1:n}}-e^{\gamma^Y_\beta-\gamma^Y_\alpha}\right)\to e^{(\gamma^Y_\beta-\gamma^Y_\alpha)}N\left(0,\sigma^2\right).$$

Since in the Dagum distribution
$\gamma^Y_\alpha=\ln\gamma_\alpha$
we have
$$\sqrt{n}\left({X_{\lfloor n\beta\rfloor+1:n}\over X_{\lfloor n\alpha\rfloor+1:n}}-{\gamma_\beta\over\gamma_\alpha}\right)\to \left({\gamma_\beta\over\gamma_\alpha}\right)N\left(0,\sigma^2\right),$$
i.e.
$$\sqrt{n}\left(r^*_{\alpha,\beta}-r_{\alpha,\beta}\right)\to r_{\alpha,\beta}N\left(0,\sigma^2\right).\eqno{(*)}$$
$\blacksquare$

\bigskip

Simple calculations show that $$\sigma^2={1\over (av)^2}
\left({1-\beta\over\beta}{1\over(1-\beta^{1\over a})^2}+{1-\alpha\over\alpha}{1\over(1-\alpha^{1\over a})^2}-2{1-\beta\over\beta}{1\over(1-\alpha^{1\over a})(1-\beta^{1\over a})}\right).
$$

Since we are interested in the estimation of the ratio $r_{\alpha,\beta}$ of the quantiles, we reparametrize the considered model. It can be seen that
$$v={\log\left({\alpha^{-1/a}-1\over \beta^{-1/a}-1}\right)\over\log r_{\alpha,\beta}}.$$ The CDF of a Dagum distribution may be written in the following form
$$F_{a, r_{\alpha,\beta}, \lambda}(x)=\left(1+\left({x\over\lambda}\right)^{-{\log\left({\alpha^{-1/a}-1\over \beta^{-1/a}-1}\right)\over\log r_{\alpha,\beta}}}\right)^{-a}$$
for $x>0$ and  $a>0$, $r_{\alpha,\beta}>0$ and $\lambda>0$.

We have $\sigma^2=(\log r_{\alpha,\beta})^2w^2(a)$, where
$$\eqalign{
w^2(a)=&\left({1\over a\log\left({\alpha^{-1/a}-1\over \beta^{-1/a}-1}\right)}\right)^2\cr
&\hskip2em\cdot\left({1-\beta\over\beta}{1\over(1-\beta^{1\over a})^2}+{1-\alpha\over\alpha}{1\over(1-\alpha^{1\over a})^2}-2{1-\beta\over\beta}{1\over(1-\alpha^{1\over a})(1-\beta^{1\over a})}\right).\cr
}$$

Let $\delta$ be a given confidence level. From $(*)$ we have (the scale parameter $\lambda$ is omitted)
$$P_{r,a}\left\{\sqrt{n}\left|{r^*_{\alpha,\beta}-r_{\alpha,\beta}\over w(a)r_{\alpha,\beta}\log r_{\alpha,\beta}}\right|\leq u_{(1+\delta)/2}\right\}=\delta,$$
where $u_{(1+\delta)/2}$ is a quantile of $N(0,1)$ distribution.

Solving the above inequality with respect to $r_{\alpha,\beta}$ we obtain confidence interval with the ends
$${r^*_{\alpha,\beta}z_{\pm}(a)\over W\left(r^*_{\alpha,\beta}z_{\pm}(a)\exp\left(z_{\pm}(a)\right)\right)},$$
where $z_{\pm}(a)={\sqrt{n}\over u_{(1\pm\delta)/2}w(a)}$ and $W(\cdot)$ is the Lambert $W$ function (see Appendix 2).

Note that the ends of the confidence interval depend on an unknown shape parameter $a$. This parameter is a nuisance parameter and must be eliminated. There are at least two methods of eliminating such nuisance parameters: estimating or appropriate averaging. In our considerations shape parameter $a$ would be estimated. There arises the problem what estimation method should be chosen. Because theoretical considerations seems to be impossible hence a simulation study was performed.

\goodbreak
\roz{3. simulation study}

Simulation study was done for different values of ratios $r_{\alpha,\beta}$ of quantiles and shape parameter $a$ (since scale parameter is not important in the problem of ratio of quantiles estimation it was taken $\lambda=1$). We take $\alpha=0.2$, $\beta=0.8$ and nominal confidence level equal to $0.95$.

Among various  methods of parameter estimation for the Dagum distribution (Dey et al. 2017) two methods were chosen. The first one is the classical method of moments (MM). In this method theoretical moments of the distribution are compared with the empirical ones. Estimators obtained by this method are solutions of the following system of equations
$$\lambda^m{\Gamma\left(a+{m\over v}\right)\Gamma\left(1-{m\over v}\right)\over\Gamma\left(a\right)}={1\over n}\sum_{i=1}^nx_i^m,\quad\hbox{for }m=1,2,3.$$
The left hand side is the $m^{th}$ moment of the Dagum distribution (see Appendix 1).

As the second method the probability-weighted moments (PWM) (see eg. Hosking et al. 1985; Ma³ecka and Pekasiewicz 2013; Pekasiewicz 2015) has been chosen. Probability-wei\-gh\-ted moments of the Dagum distribution are equal to (see Appendix 1)
$$E_{a,v,\lambda}\left[XF_{a,v,\lambda}^m(X)\right]=\lambda{\Gamma\left((m+1)a+{1\over v}\right)\Gamma\left(1-{1\over v}\right)\over(m+1)\Gamma\left((m+1)a\right)}, \quad\hbox{for }m\geq0.$$
Estimators obtained by that method are the solutions of the following system of equations (for $m=0,1,2$)
$$\cases{
{\lambda\Gamma\left(a+{1\over v}\right)\Gamma\left(1-{1\over v}\right)\over\Gamma\left(a\right)}={1\over n}\sum_{i=1}^nx_{i:n},\cr
{\lambda\Gamma\left(2a+{1\over v}\right)\Gamma\left(1-{1\over v}\right)\over2\Gamma\left(2a\right)}={1\over n}\sum_{i=1}^n{(i-1)\over(n-1)}x_{i:n},\cr
{\lambda\Gamma\left(3a+{1\over v}\right)\Gamma\left(1-{1\over v}\right)\over3\Gamma\left(3a\right)}={1\over n}\sum_{i=1}^n{(i-1)(i-2)\over(n-1)(n-2)}x_{i:n}.\cr
}$$

Estimated coverage probabilities based on $10000$ repetitions of samples of size $n=1000$ are given in Table 1 (MM) and Table 3 (PWM). In Table 2 (MM) and in Table 4 (PWM) average lengths of confidence intervals are given.

\bigskip
\setbox101=\vbox{\tabskip1em\offinterlineskip\halign to0.48\hsize{
\strut\hfil$#$\hfil&#\vrule&&\hfil$#$\hfil\cr
\multispan{7}{\bf Table 1.} Coverage probability\hfil\cr\noalign{\vskip2pt}
&&\multispan5\hfil$r_{\alpha,\beta}$\hfil\cr
a&&1.2&&1.6&&2\cr\noalign{\hrule}
0.1&&0.9493&&0.9492&&0.9494\cr
0.5&&0.9497&&0.9500&&0.9530\cr
1.0&&0.9501&&0.9518&&0.9558\cr
1.5&&0.9491&&0.9496&&0.9549\cr
2.0&&0.9475&&0.9477&&0.9492\cr
}}

\setbox102=\vbox{\tabskip1em\offinterlineskip\halign to0.48\hsize{
\strut\hfil$#$\hfil&#\vrule&&\hfil$#$\hfil\cr
\multispan{7}{\bf Table 2.} Average length\hfil\cr\noalign{\vskip2pt}
&&\multispan5\hfil$r_{\alpha,\beta}$\hfil\cr
a&&1.2&&1.6&&2\cr\noalign{\hrule}
0.1&&0.03793&&0.13175&&0.24484\cr
0.5&&0.03205&&0.11137&&0.20909\cr
1.0&&0.03025&&0.10541&&0.20010\cr
1.5&&0.03014&&0.10457&&0.19901\cr
2.0&&0.03023&&0.10442&&0.19688\cr
}}

\setbox103=\vbox{\tabskip1em\offinterlineskip\halign to0.48\hsize{
\strut\hfil$#$\hfil&#\vrule&&\hfil$#$\hfil\cr
\multispan{7}{\bf Table 3.} Coverage probability\hfil\cr\noalign{\vskip2pt}
&&\multispan5\hfil$r_{\alpha,\beta}$\hfil\cr
a&&1.2&&1.6&&2\cr\noalign{\hrule}
0.1&&0.9496&&0.9494&&0.9494\cr
0.5&&0.9496&&0.9491&&0.9490\cr
1.0&&0.9495&&0.9497&&0.9492\cr
1.5&&0.9484&&0.9481&&0.9486\cr
2.0&&0.9479&&0.9477&&0.9483\cr
}}

\setbox104=\vbox{\tabskip1em\offinterlineskip\halign to0.48\hsize{
\strut\hfil$#$\hfil&#\vrule&&\hfil$#$\hfil\cr
\multispan{7}{\bf Table 4.} Average length\hfil\cr\noalign{\vskip2pt}
&&\multispan5\hfil$r_{\alpha,\beta}$\hfil\cr
a&&1.2&&1.6&&2\cr\noalign{\hrule}
0.1&&0.03803&&0.13182&&0.24483\cr
0.5&&0.03204&&0.11077&&0.20529\cr
1.0&&0.03020&&0.10433&&0.19326\cr
1.5&&0.03009&&0.10393&&0.19249\cr
2.0&&0.03021&&0.10435&&0.19325\cr
}}

\bigskip
\centerline{\copy101\hss\copy102}
\bigskip
\centerline{\copy103\hss\copy104}
\bigskip

Since in real samples usually comprise many thousands of observations (c.f. Section 4) it was decided to use in simulations samples of size $1000$. It appears that such a size may be treated as large enough to do asymptotics: the simulated coverage probability is very close to the nominal confidence level. Of course, for larger sample sizes the coverage probability should be almost equal to the assumed confidence level.

It can also be noticed that whatever method of estimation (method of moments or of probabili\-ty\=weighted moments) is applied, probability of covering the true value of the quintile share ratio is near the nominal confidence level. It is also seen that the lengths of obtained confidence intervals are similar; it may be concluded that the length does not depend on the applied method of estimation.

It is worth noting that the method of probability-weighted moments has an advantage over the classical method of moments. Namely, the method of moments is applicable for the distributions which have at least three moments, while method of probability-weighted moments can be applied for the distributions which have at least the expected value (and thus have heavier tails). In light of the presented results of simulations, in the construction of confidence interval for quintile share ratio the method of probability-weighted moments may be recommended in estimation of shape parameter $a$ of the Dagum distribution.


\roz{4. an example of application}

In this section we present the application of the inequality of income distribution measure based on the first and the fourth quintile, i.e. $r_{0.2,0.8}$,  to income inequality analysis in Poland. Calculations are based on the sample coming form the Household Budget Survey (HBS) $2015$ provided by the Polish Central Statistical Office and being the main source of information on income and expenditure of the population of households.
\input epsf

\midinsert
\begingroup
\baselineskip=\interlinia
\vbox{
\epsfxsize=70mm
\centerline{\epsffile{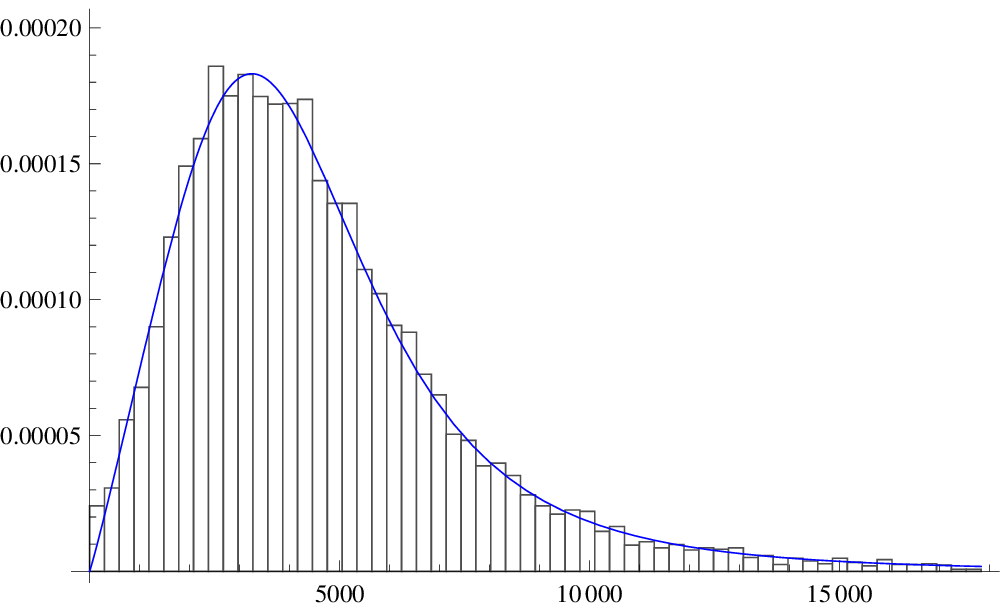}}
{\bf Figure 1.} Income distribution in Poland and fitted Dagum distribution ($a=0.6396$, $v=3.2403$, $\lambda=4961.36$)
}
\endgroup
\endinsert

The sample of size $n=13420$ was drawn. Firstly it was checked whether Dagum distribution fits data. In Figure 1 it is shown the histogram of collected data along with the fitted Dagum distribution (the probability-weighted moments method was applied). The $p$-value of the standard Kolmogorov-Smirnov test equals $0.8983$. Hence it may be accepted that sample follows Dagum distribution.

The sample quintile share ratio $r^*_{0.2,0.8}$ is $2.7600$. Application of the formula $(*)$ gives the confidence interval $(2.7081, 2.8160)$ for the population quintile share ratio $r_{0.2,0.8}$ (confidence level equals $0.95$). It may be concluded that the income in Poland is quite homogeneous, i.e. the poorest among the richest is about $2.76$ (at least $2.71$ but at most $2.82$) reacher then the richest among the poorest.

\roz{5. conclusions}

The main goal of the paper was to construct a confidence  interval for the ratio of quantiles of the Dagum distribution. According to the best knowledge of the Authors such confidence interval was never constructed. The confidence interval we propose is asymptotic. The first reason of such approach is lack of finite sample results on the distribution of the ratio of sample quantiles of the Dagum distribution. Unfortunately, the distribution of the ratio of sample quantiles derived by Maswadah (2013) was found to be analytically untractable. The second reason of considering asymptotics was that in practise the samples of income are really of large sizes. In a short simulation study it has been shown that sample size of  $1000$ may be treated as large enough to do asymptotics.

The ends of the obtained asymptotic confidence interval depend on shape parameter $a$ of the Dagum distribution. This parameter should be estimated from a sample. In a simulation study two estimators of this parameter were applied. Both estimators gave similar results.

It will be interesting to check whether the length of the confidence interval depends on the choice of the estimation method (Maximum Likelihood, Method of $L$-Moments, Method of Maximum Product of Spacings and other) of the shape parameter $a$. Theoretical solutions seem unavailable, so simulation studies are needed. Such studies are in preparation and will be  published separately.

The confidence interval constructed above is symmetrical in the following sense: the risks of underestimation and overestimation are the same. It may also be interesting to consider a~problem of constructing the shortest confidence interval. The idea of building such intervals is explained in detail in Zieliñski (2010, 2017).

\roz{acknowledgements}

The authors would like to thank Professor Anna Dembiñska (Warsaw University of Technology) for helpful discussion.

\roz{References}

\def\art#1#2#3#4#5{\noindent\hangindent=0.5truecm \hangafter=1 {\kapitaliki #1}\ (#2):\ ``#3,''\ {\it #4}, #5.\par}
\def\book#1#2#3#4{\noindent\hangindent=0.5truecm \hangafter=1 {\kapitaliki #1}\ (#2):\ ``#3,''\ {#4}.\par}

\art{Dagum, C.}{1977}{A New Model of Personal Income Distribution: Specification and Estimation}{Economie Appliquee}{30, 413-437}

\book{David, H. A. and Nagaraja, H. N.}{2003}{Order Statistics, Third Edition}{John Wiley \& Sons, Inc}

\art{Davidson, R.}{2009}{Reliable Inference for the Gini Index}{Journal of Econometrics}{150, 30-40}

\art{Dey, S., Al-Zahrani, B. and Basloom, S.}{2017}{Dagum Distribution: Properties and Different Methods of Estimation}{International Journal of Statistics and Probability}{6, 74-92, doi:10.5539/ijsp.v6n2p74}


\art{Domañski, Cz. and Jêdrzejczak, A.}{1998}{Maximum Likelihood Estimation of the Dagum Model Parameters}{International Advances in Economic Research}{4, 243-252}

\book{Encyclopedia}{2006}{Encyclopedia of Statistical Sciences, Second Edition, Volume 5}{John Wiley \& Sons, Inc}

\noindent\hangindent=0.5truecm \hangafter=1 {\kapitaliki Eurostat}\ (2016):\ ``The Eurostat regional yearbook ISBN: 978-92-79-60090-6, ISSN: 2363\=1716, doi: 10.2785/29084, cat. number: KS-HA-16-001-EN-N. (http://ec.europa.eu\hfill\break/eurostat /statistics-explained/index.php/Glossary:Income$\_$quintile$\_$share$\_$ratio).''\par

\book{Greene, W. H.}{2003}{Econometric Analysis (5th ed.)}{Prentice Hall}

\art{Hosking,  J. R. M., Wallis, J. R. and Wood, E. F.}{1985}{Estimation of the Generalized Extreme-Value Distribution by the Method of Probability-Weighted Moments}{Technometrics}{27, 251-261}

\art{Jêdrzejczak, A.}{1994}{Application of Dagum Coefficients in Investigating Income Inequalities in Poland}{Statistical Review}{41, 55-66, (in polish)}




\art{Kleiber, Ch.}{2008}{A Guide to the Dagum Distributions}{in: Modeling Income Distributions and Lorenz Curves}{Springer}


\art{Ma³ecka, M. and Pekasiewicz, D.}{2013}{A Modification of the Probability Weighted Method of Moments and its Application  to Estimate the Financial Return Distribution Tail}{Statistics in Transition}{14, 495-506}

\art{Maswadah, M.}{2013}{On the Product and Ratio of Two Generalized Order Statistics from the Generalized Burr Type-II Distribution}{Journal of Mathematics and Statistics}{9, 129-136}


\book{Pekasiewicz, D.}{2015}{Order Statistics in Estimation Procedures and Their Applications in Socio-economic Research}{University of Lodz (in polish)}

\book{Serfling, R. J.}{1980}{Approximation Theorems of Mathematical Statistics}{John Wiley \& Sons}




\art{Zieliñski, W.}{2010}{The Shortest Clopper-Pearson Confidence Interval for Binomial Probability}{Communications in Statistics - Simulation and Computation}{39, 188-193, doi: 10.1080/03610910903391270}

\art{Zieliñski, W.}{2017}{The Shortest Clopper-Pearson Randomized Confidence Interval for Binomial Probability}{REVSTAT-Statistical Journal}{15, 141-153}

\bigskip


\roz{appendix 1}

Random variable $X$ follows the Dagum distribution with parameters  $a,v,\lambda$ if its probability density function is given by the formula:
$$f_{a, v, \lambda}(x)={av\over\lambda}\left({x\over\lambda}\right)^{av-1}\left(1+\left({x\over\lambda}\right)^v\right)^{-a-1}\hbox{ for }x>0.$$
Parameters $a,v,\lambda$ are positive reals. Parameters $a$ and $v$ are shape parameters and $\lambda$ is a~scale parameter.

The distribution is unimodal if $av>1$. Otherwise it is non-modal. If $av>1$ the mode value is equal to
$$\lambda\left({av-1\over v+1}\right)^{1\over v}.$$

Moments of the random variable $X$ equal
$$E_{a,v,\lambda}X^m=\lambda^m{\Gamma\left(1 - {m\over v}\right)\Gamma\left(a + {m\over v}\right)\over\Gamma\left(a\right)},\quad\hbox{for $m<v$}.$$
Empirical moment from a sample $X_1,\ldots,X_n$, i.e.
$${1\over n}\sum_{i=1}^nX_i^m$$
is the unbiased estimator of $m^{th}$ moment of the random variable $X$.

Coefficient of skewness is equal to (for $v>3$)
$$\scriptstyle{\Gamma^2\left(a\right)\Gamma\left(a+{3\over v}\right)\Gamma\left(1-{3\over v}\right)-3\Gamma\left(a\right)\Gamma\left(a+{1\over v}\right)\Gamma\left(a+{2\over v}\right)\Gamma\left(1-{2\over v}\right)\Gamma\left(1-{1\over v}\right)+2\Gamma^3\left(a+{1\over v}\right)\Gamma^3\left(1-{1\over v}\right)
\over
\left(\Gamma\left(a\right)\Gamma\left(a+{2\over v}\right)\Gamma\left(1-{2\over v}\right)-\Gamma^2\left(a+{1\over v}\right)\Gamma^2\left(1-{1\over v}\right)\right)^{3/2}}$$
and its kurtosis (for $v>4$) is
$$\scriptstyle
{\Gamma^2\left(a\right)\left(\Gamma\left(a\right) \Gamma\left(a+{4\over v}\right)\Gamma\left(1-{4\over v}\right) +
3 \Gamma^2\left(a+{2\over v}\right) \Gamma^2\left(1-{2\over v}\right) -
4 \Gamma\left(a+{1\over v}\right) \Gamma\left(a+{3\over v}\right) \Gamma\left(1-{3\over v}\right)\Gamma\left(1-{1\over v}\right)\right)
\over
\left(\Gamma\left(a\right)\Gamma\left(a+{2\over v}\right)\Gamma\left(1-{2\over v}\right)-\Gamma^2\left(a+{1\over v}\right)\Gamma^2\left(1-{1\over v}\right)\right)^2}.
$$
The probability-weighted moments are equal to (for $m\geq0$ and $v>1$)
$$\eqalign{
E_{a,v,\lambda}\left[XF_{a,v,\lambda}^m(X)\right]
&=\int_0^\infty x\left[\left(1+\left({x\over\lambda}\right)^{-v}\right)^{-a}\right]^m{av\over\lambda}\left({x\over\lambda}\right)^{av-1}\left(1+\left({x\over\lambda}\right)^v\right)^{-a-1}dx\cr
&=\lambda{\Gamma\left((m+1)a+{1\over v}\right)\Gamma\left(1-{1\over v}\right)\over(m+1)\Gamma\left((m+1)a\right)}.\cr
}$$
Unbiased estimators (from a sample $X_1,\ldots,X_n$) of probability-weighted moments are
$${1\over n}\sum_{i=1}^nX_{i:n}\ \hbox{(for $m=0$)}\quad\hbox{ and }\quad
{1\over n}\sum_{i=1}^n{(i-1)\cdots(i-m)\over(n-1)\cdots(n-m)}X_{i:n}\ \hbox{(for $m\geq1$),}$$
where $X_{1:n}\leq\cdots\leq X_{n:n}$ are ordered statistics (Hosking et al. 1985).


\roz{appendix 2}

Lambert function $W(\cdot)$ is defined as a solution with the respect to $t$ of the equation $$te^t=z\Rightarrow t=W(z).$$
It is seen that
$$W(z)e^{W(z)}=z \Rightarrow e^{W(z)}={z\over W(z)}\Rightarrow W(z)=\ln\left({z\over W(z)}\right)\Rightarrow z={z\over W(z)}\ln\left({z\over W(z)}\right).$$
Since the solution with respect to $r$ of the equation $r\ln r=z$ is $r={z\over W(z)}$, hence
$$
A{x-r\over r\ln r}=1\Rightarrow Ax=r(\ln r+A)\Rightarrow e^AAx=\left(re^A\right)\ln\left(re^A\right)\Rightarrow r={Ax\over W(Axe^A)}.
$$
Application of the above to the equation
$$\sqrt{n}{r^*_{\alpha,\beta}-r_{\alpha,\beta}\over w(a)r_{\alpha,\beta}\log r_{\alpha,\beta}}= u_{(1+\delta)/2}$$
gives the confidence interval for the ratio $r_{\alpha,\beta}$.

\bye

The empirical (from a sample $X_1,\ldots,X_n$) probability weighted $m^{th}$ moment is defined as
$${1\over n}\sum_{i=1}^n X_{i:n}\left[F_n(X_{i:n})\right]^m,$$
where $X_{1:n}\leq\cdots\leq X_{n:n}$ are ordered statistics and $F_n(\cdot)$ is an empirical cumulative distribution function. For classical empirical CDF
$$F_n(x)={\#\left\{1\leq i\leq n:X_i\leq x\right\}\over n}$$
we obtain
$${1\over n}\sum_{i=1}^nX_{i:n}\ \hbox{(for $m=0$)},\quad\hbox{ and }\quad
{1\over n}\sum_{i=1}^n{(i-1)\cdots(i-m)\over(n-1)\cdots(n-m)}X_{i:n}\ \hbox{(for $m\geq1$).}
$$
Note that the empirical probability weighted moments are unbiased estimators of probability weighted moments of random variable $\xi$.
$$\eqalign{
&{1\over n}\sum_{i=1}^nX_{i:n},\quad\hbox{for $m=0$}\cr
&{1\over n(n-1)}\sum_{i=1}^n(i-1)X_{i:n},\quad\hbox{for $m=1$}\cr
&{1\over n(n-1)(n-2)}\sum_{i=1}^n(i-1)(i-2)X_{i:n}.\quad\hbox{for $m=2$}\cr
}$$

\bye